\documentclass[%
reprint,
 amsmath,amssymb,
 aps,
]{revtex4-2}

\usepackage{graphicx}
\usepackage{dcolumn}
\usepackage{bm}

\usepackage{gensymb}
\usepackage{upgreek}

\begin{document}

\title{Controlled creation of point defects in 3D colloidal crystals}


\author{Max P.M. Schelling}
\author{Janne-Mieke Meijer}%

\affiliation{%
Department of Applied Physics and Science Education, Eindhoven University of Technology, P.O. Box 513, 5600 MB Eindhoven, The Netherlands
}%
\affiliation{%
Institute for Complex Molecular Systems, Eindhoven University of Technology, P.O. Box 513, 5600 MB Eindhoven, The Netherlands
}%

\date{\today}
             
\begin{abstract}{
    Crystal defects crucially influence the properties of crystalline materials and have been extensively studied. Even for the simplest type of defect---the point defect---however, basic properties such as their diffusive behavior, and their interactions, remain elusive on the atomic scale. Here we demonstrate in-situ control over the creation of isolated point defects in a 3D colloidal crystal allowing insight on a single particle level. Our system consists of thermoresponsive microgel particles embedded in a crystal of non-responsive colloids. Heating this mixed particle system triggers the shrinking of the embedded microgels, which then vacate their lattice positions creating vacancy-interstitial pairs. We use temperature-controlled confocal laser scanning microscopy to verify and visualize the formation of the point defects. In addition, by re-swelling the microgels we quantify the local lattice distortion around an interstitial defect. Our experimental model system provides a unique opportunity to shed new light on the interplay between point defects, on the mechanisms of their diffusion, on their interactions, and on collective dynamics.
    }
\end{abstract}

\maketitle

\section{Introduction} 
In any crystal, structural imperfections (defects) are thermodynamically bound to occur. Defects crucially influence the mechanical, structural and optical properties of materials and for this reason have remained extensively studied \cite{Hull2011IntroductionDislocations, Phillips2001CrystalsMicrostructures}. While defects may have desirable effects --- they may enhance the conductive properties of  semiconductors \cite{Koenraad2011SingleSemiconductors}, or may act as active sites in catalysis \cite{Xie2020DefectUtilization} --- their presence can also be undesirable, causing metal fatigue or the softening of nanocrystalline metals \cite{Schitz1998SofteningSizes}.
Of particular interest are point defects (vacancies and interstitials). While these distort the lattice only locally, they play a crucial role in bulk properties such as diffusion creep and material degradation \cite{Phillips2001CrystalsMicrostructures}. A dramatic example of this occurs in radiation shielding materials, where high energy particles generate vacancy-interstitial pairs (Frenkel pairs) that are known to greatly compromise material strength and integrity  \cite{Knaster2016MaterialsFusion, Bai2010EfficientEmission, Lu2016EnhancingAlloys}.  However, despite remarkable progress in high-resolution electron microscopy studies of defect dynamics \cite{Arakawa2007ObservationLoops, Matsukawa2007One-DimensionalMetals, Moll2013Direct-Iron, Chu2022InMetal}, the atomic details of the local lattice distortion, diffusion and collective organization of point defects remain elusive.

\begin{figure}[b]
\includegraphics{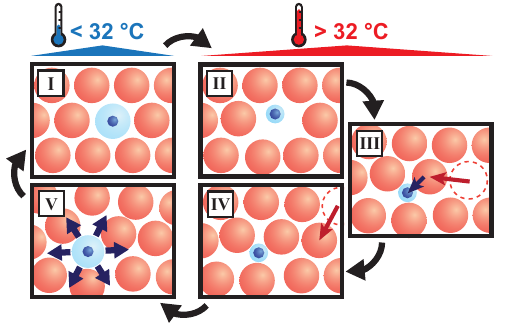}
\caption{\label{fig_1} \textbf{Controlled point defect creation.} By embedding size-tunable microgels in a crystal of non-responsive colloids, the formation of a vacancy-interstitial pair is induced. Subsequently, the lattice can be strained by re-swelling of interstitial microgels.}
\end{figure}

Here, we present an experimental platform offering a window into the basic properties and dynamics of point defects in crystalline materials using colloids. Colloids have played a central role as convenient experimental model systems to study many physical phenomena, such as phase transitions and the glass transition \cite{Gasser2001Real-spaceCrystallization, Zhang2014ExperimentalAssembly, Alsayed2005PremeltingCrystals, Peng2010MeltingFilms, Wang2016MeltingCrystals, Yunker2010ObservationTransition, Weeks2017IntroductionTransition}. Colloidal particles are small enough to experience Brownian motion, causing them to exhibit equilibrium phase behavior similar to atomic and molecular systems. Importantly, though, they are large enough to be readily observed, with single-particle resolution, using optical microscopy. For example, experiments on defects in 3D colloidal crystals have already provided unique insight into dislocations (line defects); their nucleation \cite{Schall2006VisualizingCrystals}, dynamics \cite{Schall2004VisualizationCrystals}, interactions \cite{Svetlizky2023DislocationCrystals} and width \cite{Hilhorst2011VariableSpheres}. Experiments on \textit{point} defects, however, have so far been limited to 2D systems \cite{Cao2020Pile-upCrystals}, for instance employing laser optical tweezers to artificially introduce vacancies and interstitials revealing the equilibrium configurations \cite{Pertsinidis2001EquilibriumCrystals}, diffusion \cite{Pertsinidis2001DiffusionCrystals, Kim2020DynamicalCrystal}, defect string formation \cite{Lechner2013Self-organizedCrystals} and kinetics \cite{Lechner2015EntropyCrystals, VanDerMeer2014HighlyCrystals}. So far, point defects in 3D colloidal crystals remain underexamined because experimental point defect concentrations are low \cite{Pronk2001PointCrystals} and optical tweezers cannot be used to remove or add particles in 3D crystals without distorting the full crystal \cite{Liu2016CoreShellMaterials}. 

In this Article, we overcome the challenge of the controlled generation of point defects in 3D in an experimental responsive colloidal system that breaks ground for point defect studies in 3D. We demonstrate that our system can induce both vacancies and interstitials, and we present the experimental determination of local lattice distortions around an interstitial in 3D. Our system consists of thermoresponsive poly(\textit{N}-isopropyl acrylamide) (PNIPAM) microgels embedded in a crystal of non-responsive latex colloids and is schematically depicted in Fig. \ref{fig_1}. The size of PNIPAM microgels is reversibly tunable with temperature. At low temperatures the microgels are swollen with water, while above the Volume Phase Transition Temperature (VPTT, around 32 \degree C) the microgels collapse \cite{Pelton1986PreparationN-isopropylacrylamide, Bae1990TemperatureWater} resulting in a change in their volume fraction that has been extensively studied in bulk \cite{Karg2019NanogelsTrends, Yunker2014PhysicsParticles, Wang2016MeltingCrystals}. In our system, by employing a low concentration of microgels, they will take up a lattice position in a non-responsive crystal at low temperatures. Heating the system across the VPTT leads to a significant decrease of the microgel size and allows them to migrate to an interstitial lattice site, which effectively results in the formation of a vacancy-interstitial pair. Subsequently, by lowering the temperature, re-swelling of a microgel that is located on an interstitial lattice site results in a significant local strain that we can visualize on a single-particle level. We measured the resulting lattice distortion around these interstitials and found an anisotropic strain with a decay that is comparable to results from theory \cite{Phillips2001CrystalsMicrostructures, Hall1957DistortionCrystals} and simulations \cite{VanDerMeer2017DiffusionCrystals, Alkemade2021PointColloids}. 

\section{Results and discussion} 
\begin{figure}[b]
    \includegraphics{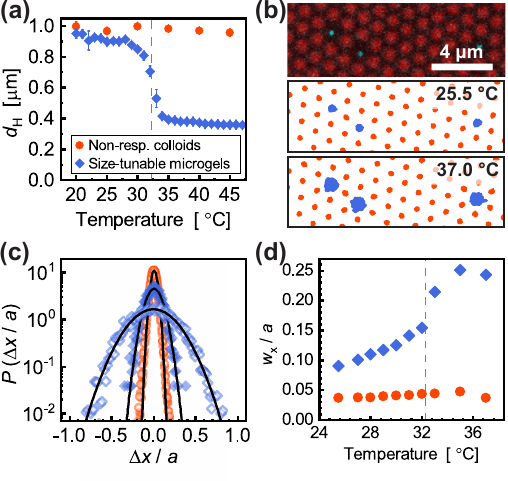}
    \caption{\label{fig_2} \textbf{Realization of the mixed system.} \textbf{(a)} Hydrodynamic diameter $d_\textrm{H}$ as a function of temperature for both non-responsive PTFEMA colloids and size-tunable PNIPAM microgels. The dashed line denotes the VPTT of the microgels. Error bars represent one standard deviation. \textbf{(b)} CLSM image of the mixed crystal at 25.5 \degree C (upper panel), and 2D trajectories (30 s, 15 fps) in the same field-of-view at 25.5 \degree C (middle panel) and 37.0 \degree C (lower panel). \textbf{(c)} Self-part of the 1D Van Hove function for reduced displacements $\Delta x / a$, where $a$ is the nearest neighbor distance ($a = 1.09 ~\upmu \mathrm{m}$) and lag time $\tau = 1 ~\mathrm{s}$. Solid symbols are at 25.5 \degree C and open symbols are at 37.0 \degree C. Solid lines are Gaussian fits using Eq. (1). \textbf{(d)} Reduced width $w_\mathrm{x} / a$ of the 1D Van Hove functions as function of temperature. Dashed line represents the VPTT from DLS.}

    \makeatletter
    \let\save@currentlabel\@currentlabel
    \edef\@currentlabel{\save@currentlabel(a)}\label{fig_2a}
    \edef\@currentlabel{\save@currentlabel(b)}\label{fig_2b}
    \edef\@currentlabel{\save@currentlabel(c)}\label{fig_2c}
    \edef\@currentlabel{\save@currentlabel(d)}\label{fig_2d}
    \makeatother
\end{figure}

\textbf{Colloidal model system.} To realize the system presented in Fig. \ref{fig_1}, two types of colloidal particles were synthesized. The microgels consist of a fluorescent, non-responsive core of poly(2,2,2-trifluoro ethyl methacrylate) (PTFEMA) and a crosslinked PNIPAM microgel shell \cite{Appel2015TemperatureSuspensions}. Fluorescent latex particles consisting of PTFEMA and coated with a thin layer of poly(oligo-ethylene glycol methacrylate) were used as non-responsive colloids (see Methods and Supplementary Information for details about the synthesis). Fig. \ref{fig_2a} shows the hydrodynamic diameter of the particles versus temperature as determined with dynamic light scattering (DLS). The VPTT of the microgels is determined as 32.3 \degree C (see Supplementary Fig. 1). Below the VPTT (20.0 \degree C), both particles have a similar size $d_\textrm{H} \approx 1.0 ~\mathrm{\upmu m}$, ensuring that the microgels can replace a PTFEMA particle in the crystal. Above the VPTT, the microgels decrease significantly in size to $d_\textrm{H}$ = $ 0.38 \pm 0.01 ~\mathrm{\upmu m}$ (37.0 \degree C) making them small enough to occupy an octahedral site of $0.41 ~\mathrm{\upmu m}$ in a close-packed crystal of $1.0 ~\mathrm{\upmu m}$ colloids.

The mixed system consists of a small fraction of microgels (0.5 - 1 \%) in a crystal of non-responsive colloids with volume fraction $0.57 \pm 0.04$ (see Methods for details). The colloids adopt a random-hexagonal close packed (\textit{rhcp}) structure upon sedimentation, which is essentially a mixture of face-centered cubic (\textit{fcc}) and hexagonal close packed (\textit{hcp}) crystal structures.  Fig. \ref{fig_2b} (upper panel) shows a 2D confocal laser scanning microscopy (CLSM) image of the mixed crystal, where the hexagonal \textit{rhcp}  planes are parallel to the flat substrate. Here, the microgels clearly take up a single lattice site, although only the fluorescent microgel cores are visible, since the shells are not fluorescently labeled. To change the microgel size in the mixed crystal, temperature-controlled CLSM experiments were performed in 2D and 3D using a VAHEAT (Interherence) temperature controller \cite{Icha2022PreciseVAHEAT}. Particle positions were obtained using Trackpy \cite{Allan2021Soft-matter/trackpy:v0.5.0} based on the Crocker-Grier centroid-finding algorithm \cite{Crocker1996MethodsStudies}.

To confirm that temperature changes only influence the microgel size and not affect the colloidal crystal structure or dynamics, we studied the mixed system behavior between 25.5 \degree C and 37.0 \degree C. Fig. \ref{fig_2b} shows 2D trajectories of both particles at 25.5 \degree C (middle panel) and 37.0 \degree C (lower panel). The trajectories clearly show that the microgels, due to their smaller size, are more mobile on their lattice site above the VPTT. In addition, we did not observe any change in the overall crystal lattice positions. 

To quantify the particle dynamics, we calculate the self-part of the Van Hove correlation function from the particle trajectories, which gives the probability distribution $P(\Delta x, \tau)$ of particle displacements $\Delta x$ along the $x$ coordinate within a lag time $\tau$. Fig. \ref{fig_2c} shows the probability distributions for $\tau = 1 ~\mathrm{s}$ for both particles (the width of the distributions does not increase for larger $\tau$; see Supplementary Fig. 2). Only for the microgels the Van Hove function shows a significant broadening with increasing temperature. The width $w_{\mathrm{x}}$ of the distributions at temperatures around the VPTT was obtained by fitting

\begin{equation}
    P(\Delta x, \tau) = \frac{1}{w_{\mathrm{x}}\sqrt{2 \pi}} \exp \left(-\frac{\Delta x^2}{2 w_{\mathrm{x}}^{2}}\right).
\end{equation}
Fig. \ref{fig_2d} shows $w_{\mathrm{x}}$ for both the PTFEMA colloids and PNIPAM microgels at each investigated temperature. As expected, no substantial change in $w_{\mathrm{x}}$ was observed for the non-responsive colloids. The sharp increase in $w_{\mathrm{x}}$ for the PNIPAM microgels is a result of their collapse around the VPTT. The size change of the microgels clearly still happens in a similar temperature regime in the dense colloidal crystal of PTFEMA particles as under dilute conditions (see DLS measurements in Fig. \ref{fig_2a}). These results confirm that by controlling the temperature we can finely tune the size of the embedded microgels without affecting the rest of the crystal.

\begin{figure}[t]
\includegraphics{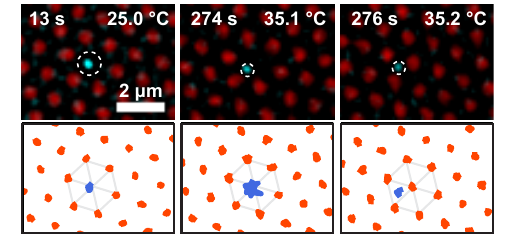}
\caption{\label{fig_3} \textbf{Formation of vacancy-interstitial pairs.} Upper panels display bandpassed CLSM images showing the formation of a vancancy-interstitial pair induced by heating across the VPTT (2.4 \degree C/min after 20 s equilibration at 25 \degree C). The vacancy is formed in an adjacent layer. Dashed circles are an indication of the microgel size. Lower panels show the corresponding trajectories (1 s).}
\end{figure}

\textbf{Formation of vacancy-interstitial pairs.} 
To create point defects, we increased the temperature of mixed colloidal crystal above the VPTT slowly. Over time, the collapsed microgels are able to spontaneously move from a regular lattice site to an interstitial position, forming the expected vacancy-interstitial pair. Fig. \ref{fig_3} shows CLSM images (upper panels) and 2D particle trajectories (lower panels) of this observed vacancy-interstitial pair formation. At the start at low temperature, the microgel is clearly embedded in the non-responsive crystal with reduced mobility. After heating the crystal to above the VPTT, the microgel starts to rattle in its cage. Since the microgel collapse leads to a reduced local pressure (similar to the case of a true vacancy \cite{Lin2016MeasuringCrystals}) at a certain moment a colloid from an adjacent layer 'hops' and takes up the original lattice site, thereby trapping the microgel in an interstitial site. We found that the formation of a vacancy-interstitial pair can happen in two ways. The first mechanism is that two spontaneous movements of the microgel and non-responsive particle occur at the same time, as shown in Fig. \ref{fig_3}. The second mechanism is the diffusion of the collapsed microgel to an octahedral interstitial site further away by passing between two non-responsive colloids, leaving behind a vacancy in the originally occupied lattice site, which we observed less frequently. For both processes there is an associated energy barrier that depends on the density of the crystal and the size of the collapsed microgel relative to the other colloids. In addition, we observed that after the creation of the vacancy-interstitial pair, the pair can split up by diffusion, resulting in isolated vacancies and interstitials which allows us to study these defects separately as well. 

The vacancy-interstitial pairs induced in this colloidal system are analogous to Frenkel pairs found in atomic systems. In metals, this type of defect does generally not occur under normal conditions due to the high formation energy of interstitials, but is predominantly generated during particle irradiation. In ionic solids Frenkel defects can be formed spontaneously because there is usually a significant size difference between the cation and anion. For the same reason, in the system presented here, only collapsed microgels are small enough to move to an interstitial site without significantly distorting the lattice.

\begin{figure*}[t]
    \includegraphics{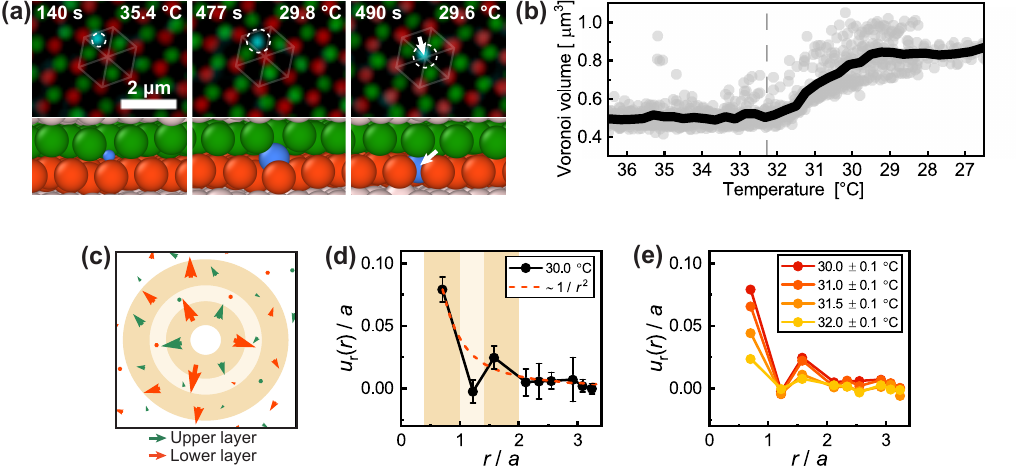}
    \caption{\label{fig_4} \textbf{Lattice strain due to interstitial swelling.} \textbf{(a)} Upper panels show an overlay of bandpassed CLSM images of the layers above (green) and below (red) an interstitial microgel (dashed circle), while the lower panels show 3D renderings at the same points in time. Upon cooling across the VPTT, the interstitial microgels (left panels) induce a local distortion on the crystal lattice (middle panels). A relaxation event involving a nearby defect causes the (nearly-)swollen microgel to take up a regular lattice site (right panels). \textbf{(b)} Voronoi cell volume of interstitial microgels while cooling across the VPTT (dashed line). Gray points are obtained from separate measurements of 14 microgels; the black line is a moving average. \textbf{(c)} Particle displacements of the layer above and below a swelling interstitial microgel at $30.0 \pm 0.1$ \degree C, measured relative to the average positions in the range 37--35 \degree C. The length of the arrows are scaled by a factor 20.  \textbf{(d)} Reduced radial component of the displacements $u_{\mathrm{r}} / a$ around an interstitial microgel as function of the distance $r / a$ from the microgel (average of five microgels), where $a$ is the nearest neighbor distance ($a \approx 1.09 ~\upmu \mathrm{m}$). Error bars denote 95\% confidence intervals obtained from the standard error. The first three layers are highlighted and correspond to the rings shown in (c). \textbf{(e)} Radial component of the displacements measured at different temperatures (an average of at least five microgels). Data obtained at $30.0 \pm 0.1$ \degree C is the same as in (d). Error bars are omitted but are typically of the same magnitude as in (d).}

    \makeatletter
    \let\save@currentlabel\@currentlabel
    \edef\@currentlabel{\save@currentlabel(a)}\label{fig_4a}
    \edef\@currentlabel{\save@currentlabel(b)}\label{fig_4b}
    \edef\@currentlabel{\save@currentlabel(c)}\label{fig_4c}
    \edef\@currentlabel{\save@currentlabel(d)}\label{fig_4d}
    \edef\@currentlabel{\save@currentlabel(d)}\label{fig_4e}
    \makeatother
\end{figure*}

\textbf{Inducing strain with interstitial microgels.} 
In our colloidal system we have the additional control that, once a collapsed microgel is on an interstitial site, lowering the temperature across the VPTT will result in re-swelling of the microgel. This local swelling provides the opportunity to apply a stress around the interstitial site and to visualize how this distorts the local crystal structure, a process that lies at the core of interstial interactions and agglomeration. To investigate the local strain due to re-swelling of interstitial microgels, we slowly cooled a mixed system with interstitial microgels across the VPTT, while simultaneously imaging the system in 3D with CLSM. Fig. \ref{fig_4a} shows an overlay of CLSM images of an interstitial microgel and the hexagonal planes directly above and below the microgel (in a pure \textit{fcc} crystal these are the (111) planes). 
For clarity, a 3D reconstruction \cite{Stukowski2010VisualizationTool} using the tracked particle positions is given below the CLSM images. The images show that cooling across the VPTT results indeed in the swelling of the microgel and a distortion of the local lattice. We note that this distortion is relatively small and therefore difficult to observe by merely looking at the CLSM images. At a certain point, a stress relaxation event occurs that results in a jump of the (nearly-) swollen microgel to a regular lattice site, pushing away the colloid that originally occupies this site. These stress relaxations happen so fast that we have not been able to fully visualize the event in 3D with the CLSM with high temporal resolution (our fastest frame rate was 0.23 frames per second for a 256x256x101 image stack). The pushed particles appear to move toward the disordered layer of colloids at the substrate, which is comparable to how in atomic systems disordered interfaces such as grain boundaries act as defect sinks.  

To determine the exact temperature range in which the buildup of lattice strain takes place, we first analyzed the Voronoi cell volume of the interstitial microgel as function of temperature \cite{Ramasubramani2020Freud:Data}. Fig. \ref{fig_4b} shows the Voronoi cell volume and the average value during the cooling process for 14 interstitial microgels. We used cooling rates of -1.0 and -3.2 \degree C/min for the experiments, but we do not observe a significant difference in the results (see Supplementary Fig. 3(a)). The curve shows that by decreasing temperature, the microgel swells and causes an increase in the Voronoi cell volume, indicating the interstitial microgel indeed pushes its direct nearest neighbors. The Voronoi cell volume of the microgels remains constant \textit{above} the VPTT (32.3 \degree C - dashed vertical line) and only increases once the temperature is \textit{below} the VPTT. Hence, only from this temperature onward the microgels start to significantly strain the crystal. For $<29$ \degree C, a constant Voronoi cell volume is present as the microgels typically have moved from an interstitial site to a regular crystal lattice site, with some exceptions that remained at the interstitial site (see Supplementary Fig. 3(b)). 

To quantify the strain caused by the swelling of interstitial microgels, we measured the local distortion that occurs in the surrounding crystal around a single interstitial microgel between 32 to 30 \degree C. Fig. \ref{fig_4c} shows the typical particle displacements measured around a single microgel by comparing the average non-responsive particle positions in the range 37--35 \degree C (i.e. fully collapsed microgel) to their positions at $30.0 \pm 0.1$ \degree C (nearly-swollen microgel on interstitial site). Similar to Fig. \ref{fig_4a}, the hexagonal planes above (green arrows) and below (red arrows) the interstitial microgel are shown. The arrows show that the strain is not isotropic: the displacements are strongest in the direction of the nearest neighbors of the interstitial microgel as these are pushed to accommodate the swelling microgel. Assuming a pure \textit{fcc} structure this displacement occurs along the cubic axes (or $\langle 100 \rangle$ directions). In the second shell of colloids around the interstitial (i.e. its next-nearest neighbors) the displacements are found to be negligible, since these colloids are not pushed by the nearest-neighbors. In contrast, the particles in the third shell are pushed by the nearest-neighbors and show a clear displacement. This non-uniform nature of the strain is also observed in the average radial component of the particle displacements $u_{\mathrm{r}}$ around a swelling interstitial microgel as plotted in Fig. \ref{fig_4d} (see Supplementary Information for more elaboration). Such anisotropic strain due to the crystal lattice has also been found with simulations \cite{VanDerMeer2017DiffusionCrystals, Alkemade2021PointColloids} and theory \cite{Hall1957DistortionCrystals}. We further note that the amount of distortion around an interstitial microgel can be tuned via the temperature as shown in Fig. \ref{fig_4e}. The results in Fig. \ref{fig_4d} indicate that the measured displacements around microgels at $30.0 \pm 0.1$ \degree C decay within a few lattice sites. The decay in the maxima of the radial displacement appears consistent with $1/r^{2}$, which can be rationalized using linear elasticity theory and, although an oversimplification, approximating the crystal as an isotropic continuum \cite{Phillips2001CrystalsMicrostructures}. Interestingly, simulations of hard-sphere systems \cite{VanDerMeer2017DiffusionCrystals} suggest an exponential decay of the displacements near a self-interstitial that ranges over more lattice sites than measured here. This discrepancy is likely due to the fact that in our results obtained at 30.0 \degree C, the microgels are not fully swollen and are therefore slightly smaller than the colloids in the crystal (see DLS results in Fig. \ref{fig_2a}). Moreover, it is difficult to predict the actual size of a swelling microgel on an interstitial site, as microgels have a soft and deformable character, and are able to osmotically deswell in dense systems \cite{Scotti2016TheSuspensions}. It is clear, however, that the interstitials can induce a local strain which ranges over several lattice sites into the non-responsive crystal.

In summary, we present here a colloidal model system in which the creation of point defects in 3D crystals can be controlled in-situ and visualized on a single-particle level. In this system, PNIPAM microgels are embedded into a non-responsive colloidal crystal where these are able to take up interstitial positions when heated above 32 \degree C (VPTT), resulting in the formation of vacancy-interstitial pairs that are similar to Frenkel pairs found in atomic systems. Due to the colloidal nature and unique temperature control the system allows us to visualize the exact moment of point defect creation and their diffusion. In addition, re-swelling of the microgels by lowering the temperature across the VPTT allows us to measure the local distortion induced by interstitials, which is otherwise problematic due to their rarity. We find that the local strain displays an anisotropy that is in agreement with results from theory and simulations. 
The system presented in this Article opens the possibility to gain new insight into point defect phenomena, such as defect interactions, as it offers in-situ control over the formation of both vacancies and interstitials, while their concentration can be controlled via the mixing ratio. This insight is valuable for understanding defect dynamics in atomic crystals, such as radiation shielding materials, while it could also help to adjust and improve the rational design and controlled defect engineering of complex crystals. 

\section{Methods} 
\subsection*{Materials and characterization}
\textbf{Colloidal particles:} The synthesis procedure used for the thermoresponsive composite microgels was based on the procedure reported by Appel et al. in Ref. \cite{Appel2015TemperatureSuspensions}. The composite microgels consist of a 192 $\pm$ 2 nm poly(2,2,2-trifluoroethyl methacrylate) (PTFEMA) core containing Pyrromethene 546 (BODIPY) dye, and a non-fluorescent PNIPAM microgel shell with 1.0 wt\% \textit{N},\textit{N}'-methylenebisacrylamide as crosslinker. The non-responsive colloids consist of PTFEMA and contain a Rhodamine B methacrylate \cite{Kodger2015PreciseMicroscopy} dye. The PTFEMA colloids are coated with thin shell of poly(\textit{oligo}-ethylene glycol methacrylate) to prevent aggregation with the microgels. A comprehensive description of the synthesis procedures of both particles is given in the Supplementary Information.

\textbf{Dynamic light scattering:} Dilute particle dispersions (0.01-0.1 wt\%) were prepared using MilliQ water (18.2 M$\Omega\cdot$cm). Particle sizes were determined with an Anton Paar Litesizer 500, using a scattering angle of 90{\degree} and laser wavelength of 658 nm, and using disposable polystyrene cuvettes. The mean hydrodynamic diameter and polydispersity index was obtained from the scattering data using the method of cumulants \cite{Frisken2001RevisitingData}.

\subsection*{Temperature-controlled microscopy experiments}

\textbf{Mixed dispersion and sample preparation:} Several mixed particle dispersions were prepared for the experiments. Typically, 4 mg of concentrated microgel dispersion (3.2 wt\%) was added to 0.4 g concentrated non-responsive PTFEMA colloid dispersion (40 wt\%) in an 1.5 mL Eppendorf tube. Sample cells were prepared using Grace Bio-Labs SecureSeal Hybridization Chambers (SKU: 621503) combined with VAHEAT smart substrates (standard range, 18x18x0.17 mm) with a 5x5 mm heated area. The mixed dispersions were vortexed thoroughly before filling the sample cells and sealing the cells using SecureSeal stickers. Images of a sealed sample cell are depicted in Supplementary Fig. 4.

\textbf{Equipment and analysis details:} Confocal laser scanning microscopy (CLSM) experiments were performed using a Nikon A1R HD25 microscope equipped with a 100x oil immersion objective (Nikon CFI Plan Apo VC, NA = 1.4). The non-responsive PTFEMA colloids containing Rhodamine B were visualized using a 561 nm laser and GaAsP PMT detector, and the composite microgels containing Pyrromethene 546 (BODIPY) were visualized with a 488 nm laser and GaAsP PMT detector. Precise control over the sample temperature was achieved by using a VAHEAT (Interherence) temperature control system. A resonant scanner and piezo objective nanopositioning system (Mad City Labs F200W) were used for fast 3D acquisition. For the 3D acquisition, we used a typical image size of 256x256 pixels in \textit{xy} and 101 pixels in \textit{z} (voxel size: 0.084x0.084x0.100 $\upmu \mathrm{m}$), resulting in a typical temporal resolution of 0.23 frames per second. Particle tracking was performed using Trackpy (v0.5.0) \cite{Allan2021Soft-matter/trackpy:v0.5.0}. In order to account for the distortion of the axial distances due to the mismatch of the refractive index of the immersion oil and the sample, a scaling factor of 0.85 was used \cite{Besseling2015MethodsIndex}. The nearest neighbor distance $a$ was determined using the first peak of the pair correlation function, which was found to be in the range of $a=1.08 ~\upmu \mathrm{m}$ to $a=1.10 ~\upmu \mathrm{m}$ in our experiments. Considering a particle size of $0.999 \pm 0.026 ~\upmu \mathrm{m}$ (see Supplementary Table I), we find an average volume fraction of $0.57 \pm 0.04$.

\section{Acknowledgments}
We thank C. Storm and L. Filion for fruitful discussions. We acknowledge T.F. Sparidans for performing DLS measurements. J.M.M. acknowledges financial support from the Netherlands Organization for Scientific Research (NWO) (016.Veni.192.119).

\end{document}



\title{SUPPLEMENTARY INFORMATION \\ Controlled creation of point defects in 3D colloidal crystals}

\author{Max P.M. Schelling}
\author{Janne-Mieke Meijer}%
 
\affiliation{%
 Department of Applied Physics and Science Education, Eindhoven University of Technology, P.O. Box 513, 5600 MB Eindhoven, The Netherlands
}%

\affiliation{%
 Institute for Complex Molecular Systems, Eindhoven University of Technology, P.O. Box 513, 5600 MB Eindhoven, The Netherlands
}%

\date{\today}

\maketitle 

\section{Particle synthesis and characterization}
\textbf{Particle synthesis.}
All chemicals were used as received from the suppliers. Water used in the syntheses was purified using a MilliQ\textsuperscript{\textregistered} Direct 8 system (18.2 M$\Omega\cdot$cm). The synthesis procedure used for the thermoresponsive composite microgels was based on the procedure reported in [J. Appel et al., \textit{Part. Part. Syst. Charact.}, \textbf{32}, 764–770 (2015)]. For the synthesis of the core particles of the composite microgels, the following compounds were mixed in a 100 mL round-bottom flask: 8.93 g 2,2,2-trifluoroethyl methacrylate (TFEMA, TCI Chemicals, 98\%), 0.994 g \textit{N}-isopropyl acrylamide (NIPAM, Merck, 97\%), 16.9 mg sodium dodecyl sulfate (SDS, Merck, 98.5\%), and 33 mL water. To make sure the cores are fluorescent, $\sim 10$ mg Pyrromethene 546 (Merck) was added. The mixture was heated to 75 \degree C while stirring at 400 RPM (Fisherbrand Oval PTFE Stir Bar, 25x12mm) and bubbling with $\mathrm{N_{2}}$ for 30 min. An initiator solution was prepared by dissolving 50.8 mg potassium persulfate (KPS, Merck, 99\%) in 2.5 mL water. The polymerization was initiated by injection of the KPS solution. The reaction mixture turned turbid after several minutes, indicating the formation of latex particles. After 4 hours, the heating was removed and the reaction mixture was allowed to cool down to room temperature. The particle dispersion was filtered to remove any coagulum. 

For the PNIPAM shell synthesis, 1.19 NIPAM, 16.8 mg \textit{N},\textit{N'}-methylenebisacrylamide (BIS, Merck, 99\%), 1 mL core dispersion (21.5 wt\%) and 50 mL water were brought together in a 100 mL round-bottom flask. After heating the mixture to 75 \degree C and bubbling with $\mathrm{N_{2}}$ for 30 min, the polymerization was again initiated by injection of a KPS solution (51.1 mg in 1.0 mL water). The turbidity of the mixture increased several minutes after addition of the initiator. The heating was removed after 4 hours and the resulting microgel dispersion was filtered to remove any coagulum. The microgel dispersion was cleaned by repeated centrifugation and redispersion in MilliQ water.

A similar procedure as for the synthesis of the microgel cores was used for the seeds of the non-responsive particles, without addition of NIPAM, and stirring at 600 RPM. For the seed particle synthesis, 9.98 g TFEMA, 32.0 mg SDS and 33 mL water were mixed in a 100 mL round-bottom flask.  Furthermore, a Rhodamine B methacrylate (0.2 mL, $\sim$0.1 g/mL in toluene) was used as fluorescent dye instead of Pyrromethene 546. The synthesis procedure used for Rhodamine B methacrylate was based on the procedure in [T.E. Kodger et al., \textit{Sci. Rep.}, \textbf{5}, 14635 (2015)]. The polymerization was initiated by injection of a KPS solution (52.3 mg in 2.5 mL water).

To obtain colloids that match the size of the microgels, a seeded emulsion polymerization was performed in which 10.0 g TFEMA, 0.5 mL core dispersion (20.6 wt\%) and 40 mL water were brought together in a 100 mL round-bottom flask. The mixture was heated to 75 \degree C while bubbling with $\mathrm{N_{2}}$ for 30 min and stirring at 600 RPM. The polymerization was initiated by injection of a KPS solution (100.8 mg in 3.0 mL). After 4 hours, the heating was removed and the reaction mixture was allowed to cool down to room temperature and filtered to remove any coagulum. The PTFEMA particle surfaces were covered with polyethylene glycol methyl ether methacrylate (PEGMA, Mw$\sim$500 g/mol, TCI Chemicals). 0.15 g PEGMA and 12.5 mg ethylene glycol dimethacrylate (EGDMA, Merck) were added together with water and 10.0 mL PTFEMA particle dispersion (16.7 wt\%) into a round-bottom flask. Again, the mixture was heated to 75 \degree C while bubbling with $\mathrm{N_{2}}$ for 30 min. The polymerization was  initiated by injection of a KPS solution (29.0 mg in 1.0 mL water), after which the polymerization proceeded for 4 hours while stirring at 600 RPM. The particle dispersion filtered and cleaned by repeated centrifugation and redispersion in water.
\newpage
\textbf{Characterization.} 
The hydrodynamic diameters of the particles obtained from the syntheses are given in Supplementary Table \ref{table_s1}.
The hydrodynamic size of the of the final particles (PFEMA-CS-P and CM-CS in Supplementary Table \ref{table_s1}) was measured for a range of temperatures between 20 \degree C and 50 \degree C, as shown in Fig. 2(a). The Volume Phase Transition Temperature (VPTT) of the microgels was determined by fitting the hydrodynamic diameter $d_{\mathrm{H}}$ with a Boltzmann sigmoid function

\[
    d_{\mathrm{H}}(T) = \frac{d_{\mathrm{H}}^{\mathrm{sw}} - d_{\mathrm{H}}^{\mathrm{co}}}{1 + e^{\frac{T - VPTT}{dT}}} + d_{\mathrm{H}}^{\mathrm{co}},
\]
where $T$ is the temperature, and $VPTT$, $dT$, $d_{\mathrm{H}}^{\mathrm{sw}}$ and $d_{\mathrm{H}}^{\mathrm{co}}$ are fit parameters. The result of fitting this function is given in Supplementary Fig. \ref{fig_S2}.

\begin{table}[h]
\caption{\label{table_s1}%
Hydrodynamic diameter $d_{\mathrm{H}}$ of the particles measured at 20 \degree C.}
\begin{tabular}{lcc}
\hline \hline
\textrm{Label}& Description &$d_{\mathrm{H}}$ $\pm$ \textrm{St. Dev. (nm)} \\
\colrule
CM-C       & PTFEMA cores of composite microgels & 192 $\pm$ 2 \\  
CM-CS      & Composite microgels (PTFEMA core + PNIPAM shell) & 952 $\pm$ 31 \\ 
\hline
PFEMA-C    & PTFEMA cores of non-responsive colloids & 223 $\pm$ 2 \\  
PFEMA-CS   & PTFEMA colloids (not coated) & 957 $\pm$ 24 \\ 
PFEMA-CS-P & PTFEMA colloids with PEGMA layer & 999 $\pm$ 26 \\ 
\hline \hline
\end{tabular}
\end{table}

\newpage
\section{Lattice distortion by swelling of interstitial microgels}
In Supplementary Fig. \ref{fig_S4a}, we plot the same data as in Fig. 4(b) in the main article but indicating the two cooling rates used in the experiments. The results show no difference in the onset of swelling for cooling rates of -3.2 \degree C/min and -1.0 \degree C/min. In Supplementary Fig. \ref{fig_S4b}, we highlight the measured Voronoi cell volume of two microgels (corresponsing to cooling rate -1.0 \degree C/min). The data shows a significantly different Voronoi cell volume after cooling across the VPTT (32.3 \degree C). The reason for this large difference can be explained by looking at the corresponding CLSM images in the panels in Supplementary Fig. \ref{fig_S4b}. One microgel shows a complete relaxation to a regular lattice site (resulting in a larger Voronoi cell volume), while the other remains on an interstitial site (resulting in a smaller Voronoi cell volume). Since not all microgels show a complete relaxation to a regular lattice site after cooling across the VPTT, a large spread in Voronoi cell volume is observed in Fig. 4(b) and Supplementary Fig. \ref{fig_S4a}.

In Supplementary Fig. \ref{fig_S5}, the radial displacements around a single interstitial microgel at $30.0 \pm 0.1$ \degree C is shown as example. The displacements are calculated by subtracting the non-responsive particle positions around a fully-collapsed interstitial microgel (typically $>33$ \degree C, averaged over 10--40 frames) from the particle positions positions at a specific temperature (averaged over several frames), i.e.  $30.0 \pm 0.1$ \degree C in case of Supplementary Fig. \ref{fig_S5}, and taking the radial component. We typically observe a large spread in the measured radial displacements of non-responsive colloids around the microgel (gray points), from which we calculate the average radial displacements around a single microgel (black points). For the radial displacements shown in Fig. 4(d) in the main text, we have further averaged over measurements of five interstitial microgels.

\newpage
\section{Supplementary Figures}
\begin{figure}[h]
\includegraphics[scale=1.1]{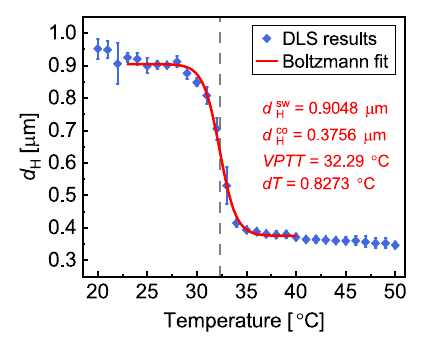}
\caption{\label{fig_S2} \textbf{Determination of the VPTT.} Boltzmann fit of the hydrodynamic diameter $d_{\mathrm{H}}$ of the microgels (CM-CS). Gray dashed line indicates the VPTT (32.3 \degree C) determined from the fit.}
\end{figure}

\begin{figure}[h]
    \includegraphics[scale=1.1]{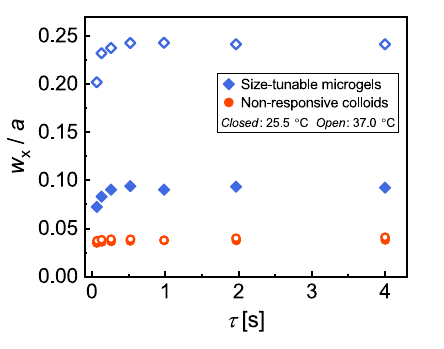}
    \caption{\label{fig_S3} \textbf{Width of the Van Hove function.} Reduced width $w_{\mathrm{x}} / a$ of the self-part of the Van Hove function as function of lag time $\tau$ for both particle types for 25.5 \degree C and 37.0 \degree C. The width $w_{\mathrm{x}}$ was obtained by fitting of Eq. (1) in the main text. Due to 'caging' of the particles on their lattice site, the width $w_{\mathrm{x}}$ remains constant for lag times $\tau>0.5$ s.}
\end{figure}

\begin{figure}[h]
    \includegraphics[scale=2]{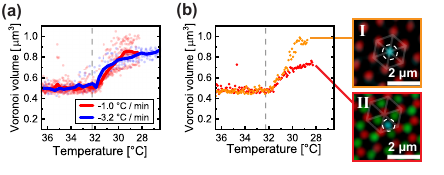}
    \caption{\label{fig_S4} \textbf{Voronoi volume of swelling interstitial microgels.} (a) Voronoi cell volume of interstitial microgels while cooling across the VPTT for two cooling rates. Data for each cooling rate consists of 7 microgels; the solid lines are moving averages. (b) The Voronoi cell volume of two microgels (-1.0 \degree C/min): one shows a complete relaxation of the microgel to a regular lattice site (see panel I), and the other an incomplete relaxation (see panel II) upon cooling. In panel I, the microgel (cyan) is located on a regular lattice site in the layer of hexagonally packed colloids (red). In panel II, the microgel (cyan) is located on an octahedral intersitial site between two hexagonally packed layers (red and green). For clarity, the brightness of the CLSM images in the panels is enhanced.}
    
    \makeatletter
    \let\save@currentlabel\@currentlabel
    \edef\@currentlabel{\save@currentlabel(a)}\label{fig_S4a}
    \edef\@currentlabel{\save@currentlabel(b)}\label{fig_S4b}
    \makeatother
\end{figure}

\begin{figure}[h]
    \includegraphics{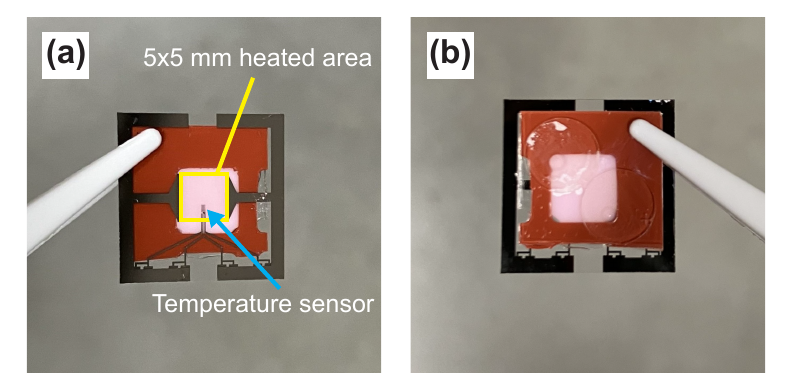}
    \caption{\label{fig_S1} \textbf{Sample cells.} A VAHEAT sample cell used in the experiments filled with a mixed particle dispersion: (a) bottom side, and (b) top side.}
\end{figure}

\begin{figure}[h]
    \includegraphics[scale=2]{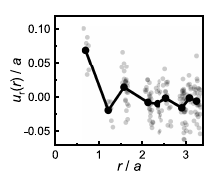}
    \caption{\label{fig_S5} \textbf{Typical displacements measured around a single interstitial microgel.} Reduced radial component of the displacements $u_{\mathrm{r}} / a$ around a single interstitial microgel as function of the distance $r / a$ from the microgel at $30.0 \pm 0.1$ \degree C, where $a$ is the nearest neighbor distance ($a \approx 1.09 ~\upmu \mathrm{m}$). The gray points are the measured displacements of individual non-responsive colloids around the microgels; the black points are the averaged displacements.}
\end{figure}